\documentclass[fleqn,usenatbib]{mnras}

\usepackage{newtxtext,newtxmath}

\usepackage[T1]{fontenc}
\usepackage{ae,aecompl}

\usepackage{graphicx}	
\usepackage{amsmath}	
\usepackage[dvipsnames]{xcolor}
\usepackage[normalem]{ulem}

\newcommand{\carito}[1]{{\color{black} #1}}

\newcommand{\blue}[1]{{\color{black} #1}}
\newcommand{\red}[1]{{\color{black} #1}}
\newcommand{\bluetwo}[1]{{\color{black} #1}}

\title[Effects of the stellar wind on the Ly-$\alpha$ transit]{Effects of the stellar wind on the Ly-alpha transit of close-in planets}

\author[S. Carolan et al.]{
S. Carolan,$^{1}$\thanks{E-mail: carolast@tcd.ie }
A. A. Vidotto,$^{1}$
C. Villarreal D'Angelo,$^{2, 1}$
G. Hazra$^{1}$
\\
$^{1}$School of Physics, Trinity College Dublin, College Green, Dublin 2, Ireland\\
$^{2}$Observatorio Astron\'omico de C\'ordoba - Universidad Nacional de C\'ordoba. Laprida 854, X5000BGR. C\'ordoba, Argentina
}

\date{Accepted XXX. Received YYY; in original form ZZZ}

\pubyear{2020}

\begin{document}
\label{firstpage}
\pagerange{\pageref{firstpage}--\pageref{lastpage}}
\maketitle

\begin{abstract}
We use 3D hydrodynamics simulations followed by synthetic line profile calculations to examine the effect increasing the strength of the stellar wind has on observed Ly-$\alpha$ transits of a Hot Jupiter (HJ) and a Warm Neptune (WN). 
We find that increasing the stellar wind mass-loss rate from 0 (no wind) to 100 times the solar mass-loss rate value causes reduced atmospheric escape in both planets (a reduction of 65\% and 40\% for the HJ and WN, respectively, compared to the `no wind' case). For weaker stellar winds \blue{(lower ram pressure)}, the reduction in planetary escape rate is very small. However, as the stellar wind becomes stronger, the interaction happens deeper in the planetary atmosphere and, once this interaction occurs below the sonic surface of the planetary outflow,  \blue{ further reduction in evaporation rates is seen}. We classify these regimes in terms of the geometry of the planetary sonic surface. \blue{ ``Closed'' refers to scenarios where the sonic surface is undisturbed, while ``open'' refers to those where the surface is disrupted.} We find that the change in stellar wind strength affects the Ly-$\alpha$ transit in a non-linear way \blue{(note that here we do not include charge-exchange processes)}. Although little change is seen in planetary escape rates ($\simeq 5.5\times 10^{11}$g/s) in the closed to partially open regimes, the Ly-$\alpha$ absorption \bluetwo{(sum of the blue [-300, -40 km/s] \& red [40, 300 km/s] wings)} changes from 21\% to 6\% as the stellar wind mass-loss rate is increased in the HJ set of simulations. For the WN simulations,  escape rates of $\simeq 6.5\times 10^{10}$g/s can cause transit absorptions that vary from 8.8\% to 3.7\%, depending on the stellar wind strength. 
We conclude that the same atmospheric escape rate can  produce a range of absorptions depending on the  \blue{stellar wind} and that neglecting  \blue{this} in the interpretation of Ly-$\alpha$ transits can lead to underestimation of planetary escape rates. 
\end{abstract}




\section{Introduction}
\label{sec:Intro}

Close-in exoplanets experience high levels of irradiation from their host stars, causing a substantial amount of photoevaporation of their atmospheres \citep{2003ApJ...598L.121L,2004A&A...419L..13B,2004Icar..170..167Y}. The amount of atmospheric escape determines the lifespan of a planet's atmosphere \citep[e.g.,][]{2015ApJ...815L..12J,Daria2020}, which is a key contributor to planetary habitability \citep{Lingam2018, Dong2018}, and is thought to shape the observed mass-radius distribution of close-in exoplanets \citep{Kurokawa2014, Owen2018, Berger2020}. \bluetwo{Additionally, atmospheric escape is believed to shape the period-radius distribution of close-in exoplanets, giving rise to the ``evaporation desert" and the ``radius valley". The evaporation desert, also known as the Neptunian desert, affects gas giants in short orbit \citep{2016A&A...589A..75M}. The radius valley is an under-population of }
exoplanets with radius between 1.5 and 2.0 Earth radii and orbital periods lower than about 100 days \citep{2013ApJ...763...12B, 2017AJ....154..109F}. More direct observational signatures of atmospheric escape have been found in transmission spectroscopic transits of a few planets, such as HD209458b \citep{VM2003}, HD189733b \citep{LDE2010, LDE2012, Jensen2012, BJB2013}, GJ436b \citep{Kulow2014, Ehrenreich2015}, GJ3470b \citep{Bourrier2018}  and some others. These observations are often done in Ly-$\alpha$ line, where absorption is a consequence of neutral hydrogen leaving the planet due to the outflow generated by high energy radiation from the host star.

Once the planetary atmosphere expands and escapes, it interacts with the stellar wind, which shapes the geometry of the escaping atmosphere. This interaction can create different structures in the escaping atmosphere, such as a comet-like tail, trailing behind the planet, and a stream of material oriented ahead of the planet's orbit and towards the star. The size and presence of these structures depend on a few key properties in the system, such as the orbital velocity, the ram pressure of the stellar wind, and tidal forces exerted by the star \citep{Matsakos2015, 2015ApJ...805...52P,Shaikhislamov2016}. For example, of the stellar wind is weak and the tidal forces are strong, a stream of planetary material can be created towards the star. If the stellar wind is strong, the orientation of the comet-like tail can become more aligned with the star-planet line. These different structures can be detected in Ly-$\alpha$, as they can contain a significant portion of neutral hydrogen. Because these structures are not spherically symmetric, they cause asymmetries in the transit light curve (e.g., an early ingress or late egress), and also affect the line profile (larger absorption in the blue and/or red wings of the line) \citep{VM2003,LDE2012,Kulow2014,Ehrenreich2015,Bourrier2018}. 

Using simulations, we can model the interactions between the stellar and planetary outflows, gaining key insights into the characteristics of the system. In the present work, we focus more specifically on the role of the stellar wind on the interaction with the escaping atmosphere and in particular whether the stellar wind can affect the amount of planetary escape. 
One crucial point in the theory of astrophysical flows, such as stellar winds and accretion disks, is the presence of ``critical points'' \citep[e.g.][]{Parker1958,weberdavis1967}. In the case of hydrodynamic outflows, such as the escaping atmospheres of close-in giants, one important critical point is the sonic point, which represents the location beyond which the planetary outflow becomes super-sonic.  
Planetary atmospheric escape models have suggested that the position of the sonic point (or surface, in the case of 3D geometries) in relation to the position where the interaction with the stellar wind happens, can act to reduce planetary escape rate \citep{Christie2016, Vidotto2020}.
It has been suggested that if the stellar wind interacts with the super-sonic escaping atmosphere, the information cannot propagate upstream and it does not affect the inner regions of the planetary outflow. However, if this interaction occurs where the escaping atmosphere is sub-sonic, the inner outflow can be altered and therefore the escape rate can be affected \citep{Vidotto2020}. In the latter scenario, the stellar wind would act to confine the planetary outflow and reduce/prevent their escape \citep{Christie2016}. 

Although an increase body of work on the 3D interaction between planetary atmospheres and stellar winds has become available recently    \citep[][Villarreal D'Angelo et al. Submitted]{Bisikalo2013,Shaikhislamov2016, Schneiter2016, CN2017, Carolina2018, McCann2019,  Khodachenko2019, Esquivel2019, Debrecht2020}, the effect of stellar wind `confinement' of planetary atmospheres has not yet been studied in 3D. For that, the interaction must occur within the sonic surface of the planetary outflow, and to the best of our knowledge, the aforementioned 3D studies have focused on the interaction that happens when the planetary outflow has already reached super-sonic speeds. 
To best model the confinement,  the planetary outflow must be launched from the surface of the planet, as the inner regions of the planetary outflow must be examined in order to accurately quantify changes in the escape rate. This requires high resolution close to the planet, in which case a planet-centric single body model is therefore preferred (`local' simulations), as opposed to `global' models that incorporate both the star and the planet in the numerical grid.

\begin{figure}
    \centering
    \includegraphics[width=\columnwidth]{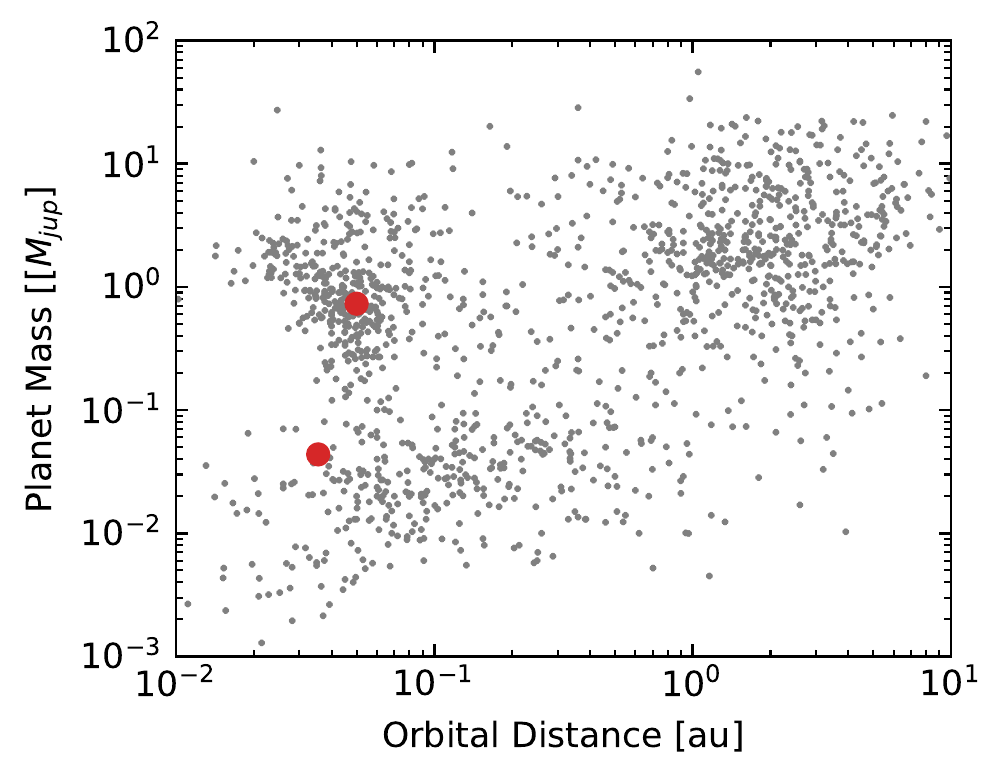}
    \caption{The masses vs orbital semi-major axis of all confirmed exoplanets are marked in grey (exoplanetarchive.ipac.caltech.edu). The red points mark the positions of the two planets examined in this work, \blue{they were chosen to have  characteristics similar to GJ3470b, and HD209458b but orbiting a more active star}.}
    \label{fig:ExoCatalogue}
\end{figure}

In this work we perform 3D local hydrodynamic simulations of atmospheric escape in close-in exoplanets, including the interaction with the stellar wind. We vary the strength of the stellar wind to investigate the effects it has on confining the outflowing atmosphere and on the atmospheric escape rate. We chose two exoplanetary systems (shown in figure \ref{fig:ExoCatalogue}, similar to HD209458b \blue{but orbiting a more active star} and GJ3470b) to represent typical close-in gas giants.  The details of our 3D model are discussed in section \ref{sec:Modelling}, with the results of these models presented in section \ref{sec:MassLoss}. We compute the synthetic observations in Ly-$\alpha$ transits in section \ref{sec:Raytrace}, and we show that properly accounting for the presence of the stellar wind can affect the interpretation of spectroscopic transits. In particular, not including the stellar wind interaction can lead to an underestimation of atmospheric escape rates detected. A discussion of our results is shown in section \ref{sec:Discussion} and we present our conclusions in section \ref{sec:Conclusion}. 

\section{3D modelling of the interaction between the stellar wind and planetary atmospheric escape}
\blue{To model the interaction between the stellar wind and the escaping atmosphere, we use two numerical setups. Firstly, we run 1D radiation hydrodynamic simulations to simulate the photoevaporation \carito{of the planetary atmosphere}. This yields an escape rate and velocity structure for a given atmosphere. These are then used to inform our 3D isothermal simulations, constraining the two free parameters: the base density and temperature of the outflowing atmosphere. Below, we detail our 1D radiation hydrodynamic model and 3D hydrodynamic model.}

\subsection{1D Model}\label{sec_1dcode}
\blue{Here, we briefly present the main characteristics of our 1D radiation hydrodynamics calculation, and refer the reader to \citet{Allan2019} or  \citet{MurrayClay2009} for further details.} This model treats the escaping atmosphere as a fluid. As well as the standard conservation of mass and momentum equations (discussed in section \ref{sec:Modelling}) this model solves an energy equation which contains additional heating/cooling terms
\blue{to consider heating from photoionisation and Ly-$\alpha$ cooling} (in the 1D model we do not include the Coriolis effect, and set $\gamma = 5/3$). 
This model also solves an equation of ionisation balance:

\begin{equation}
    \frac{ n_n F_{\rm{EUV}} e^{-\tau} \sigma_{\nu_0} }{ e_{\rm{in}} } = n_p^2 \alpha_{\rm{rec}} + \frac{1}{r^2}\frac{d}{dr}(r^2 n_p u),
\end{equation}

\noindent where $n_n$ and $n_p$ are the number densities of neutral and ionised hydrogen, $F_{\rm{EUV}}$ is the EUV flux received by the planet, $\tau$ is the optical depth to ionising photons, $\sigma_{\nu_0}$ is the cross section for the ionisation of hydrogen, and $\alpha_{\rm{rec}}$ is the radiative recombination coefficient of hydrogen ions. This model assumes that the incoming EUV flux is concentrated at an energy of $e_{\rm{in}} = 20$ eV \citep{MurrayClay2009}. 

\begin{figure}
    \centering
    \includegraphics[width=\columnwidth]{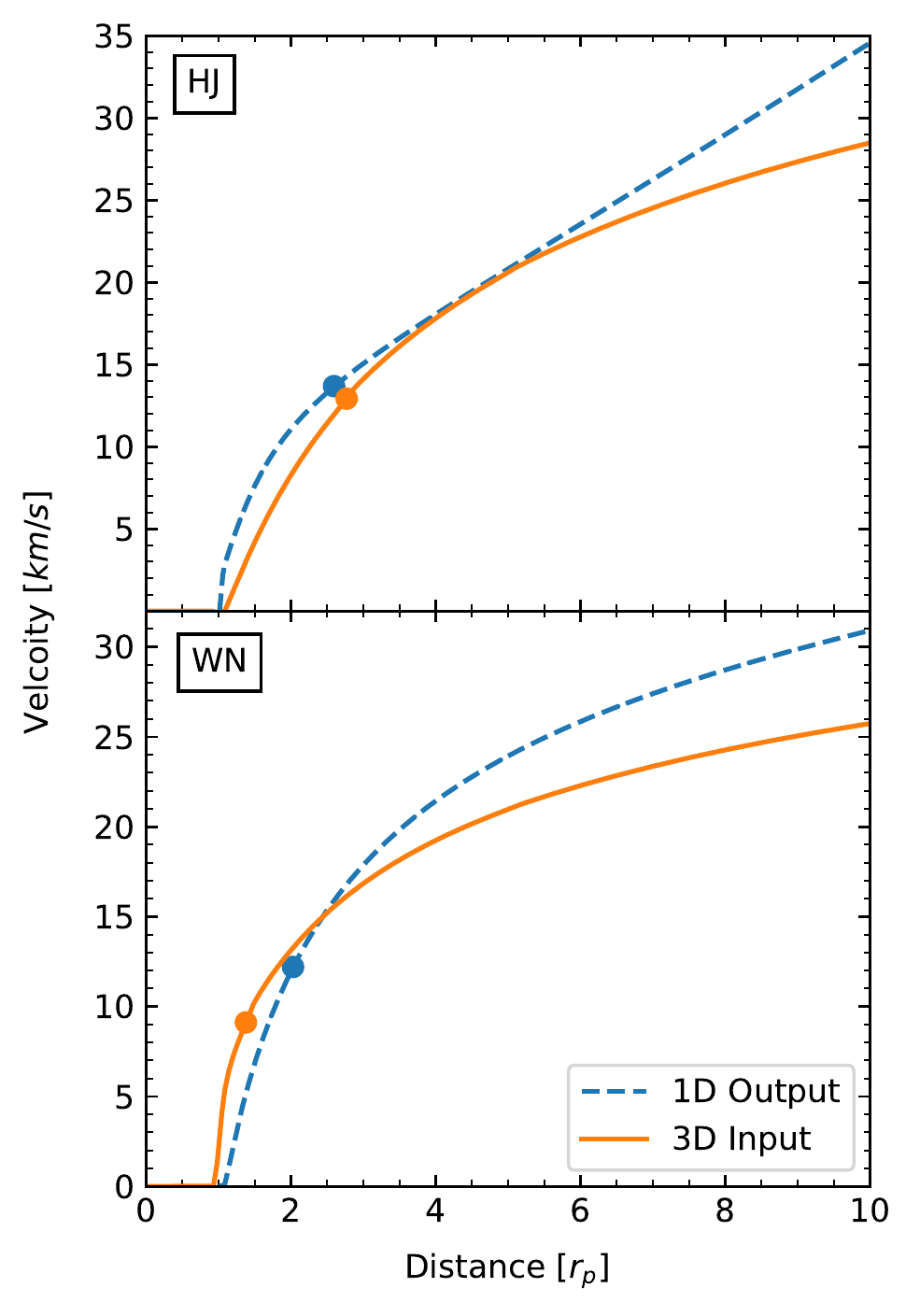}
    \caption{\blue{Velocity vs distance for the escaping atmosphere of both of our models. The dashed blue line shows the output of the 1D model, while the solid orange shows the initial condition we implement in our isothermal 3D model.  The base density in the 3D model is then adjusted so that the same escape rate is obtained in both the 1D and 3D models. The circles mark the position of the sonic point in each model. }}
    \label{fig:model_comp}
\end{figure}

This model takes the planetary parameters (table \ref{tab:planets}) as input, as well as the EUV flux from the host star: for the hot Jupiter, we used $L_{\rm{EUV, HJ}} =2.06\times10^{-5}L_\odot,$ (chosen to be 25 times larger than that assumed in \citealt{MurrayClay2009}) and for the warm Neptune, we used $L_{\rm{EUV, WN}} =  3.73 \times10^{-6} L_\odot$ (\citealt{Bourrier2018}). As a result, this yields an escape rate, as well as the velocity and ionisation fraction ($f_{\rm{ion}} = n_p/(n_p+n_n)$) as a function of distance. With this information, we can constrain the two free parameters of the 3D isothermal model: the base density and temperature \carito{at the planetary radius (inner boundary)}. A unique temperature is chosen such that the velocity structure in 3D best matches the velocity structure from the 1D model, seen in figure \ref{fig:model_comp}. Similarly, the base density in the 3D simulations is adjusted, so that it and the matched velocity structure reproduce the escape rates resultant from the 1D model ($\dot{m}_0$ in table \ref{tab:planets}). \bluetwo{By} matching \bluetwo{the} velocity structure and escape rate, we ensure that the ram pressure of the escaping atmospheres in our 3D simulations without a stellar wind \bluetwo{is} close to that of the 1D model. We can then inject a stellar wind to examine the interaction between the stellar and planetary outflows, having vastly saved computational time by informing our 3D model with the 1D model.

\begin{table*}
    \centering
    \caption{The planet and stellar properties in each set of models. $M_p$ and $r_p$ describe the planet's mass and radius, $a$ is the orbital distance. $\dot{m}_0$ is the planetary atmospheric escape rate (with no stellar wind), $u_{\rm kep}$ is the Keplerian velocity of the system.  \blue{$T_p$ and $n_{0,p}$ are the temperature and base density }of the planet's outflowing atmosphere \blue{in the 3D model that were found to best match the results of the 1D model}. $\dot{M}$, $R_*$ and $T_*$ describe the stellar mass, radius and stellar wind temperature, respectively, while the stellar wind radial velocity at each planet's orbital distance is given as $u_{\rm local}$. \bluetwo{Finally, $F_{\rm{EUV}}(a)$ is the EUV flux at the planet's orbital distance.} }
    \begin{tabular}{ccccccccccccc}
        \hline
        Planet & $M_p$ & $r_p$ & $a$ & $\dot{m}_0$ & $u_{\rm{kep}}$ & $T_p$ &$n_{0,p}$ & $M_*$ & $R_*$ & $T_*$ & $u_{\rm{local}}$ & \bluetwo{$F_{\rm{EUV}}(a)$}\\
         &  [$M_{\rm{Jup}}$] & [$R_{\rm{Jup}}$] & [au] & [$10^{10}$ g/s] & [km/s] & [$10^4$ K] &[$10^9$ cm$^{-3}$] & [$M_{\odot}$] & [$R_\odot$] & [$10^6$ K] & [km/s] & [erg/cm$^2$/s]\\
        \hline
        HJ & 0.7 & 1.4 & 0.050 & 58 & 147 & $1$ & 3.86 & 1.00 & 1.00 &$2$ & 400 & $1.12\times10^4$\\
        WN & 0.04 & 0.41 & 0.036 & 6.5& 116 & $0.5$ & 3.03 & 0.54 & 0.55 & $1$ & 240 & $3.92\times10^3$\\
        \hline
    \end{tabular}
    \label{tab:planets}
\end{table*}

\subsection{Setup of our 3D models}
\label{sec:Modelling}

To investigate the effects of the stellar wind on the reduction of atmospheric escape rates, we simulate the environments around close-in exoplanets using the Space Weather Modelling Framework \citep[SWMF, ][]{Toth-swmf}. SWMF has previously been used to study, e.g., various solar system objects \citep{Sterenborg2011, Ma2013, Jia2015, Jia2016, Carolan2019}, the solar wind \citep[e.g.][]{2004JGRA..10902107M,2011ApJS..194...23V} and stellar winds \citep[e.g.][]{Vidotto2018,2019MNRAS.485.4529K}. For our investigation, we create a new user implementation \blue{to SWMF. This involves the design of new inner and outer boundary conditions, and additional source terms in the hydrodynamic equations, all of which are outlined below.}

We model the stellar wind and escaping atmosphere as isothermal winds \citep{Parker1958}, we do not include the effects of magnetic fields (this will be the topic of a future study), and consider only hydrogen in our simulations. There are two unknowns in the Parker wind \carito{model}, the base density and temperature. To guide the selection of temperature and density of the planetary outflow, we use information from our 1D model \citep[][see section \ref{sec_1dcode}]{Allan2019}. For the stellar winds we chose temperatures appropriate for stars hosting close-in exoplanets (see table \ref{tab:planets}), and vary the base density to control the stellar mass loss rate.

\begin{figure}
    \centering

    \includegraphics[width=0.5\textwidth]{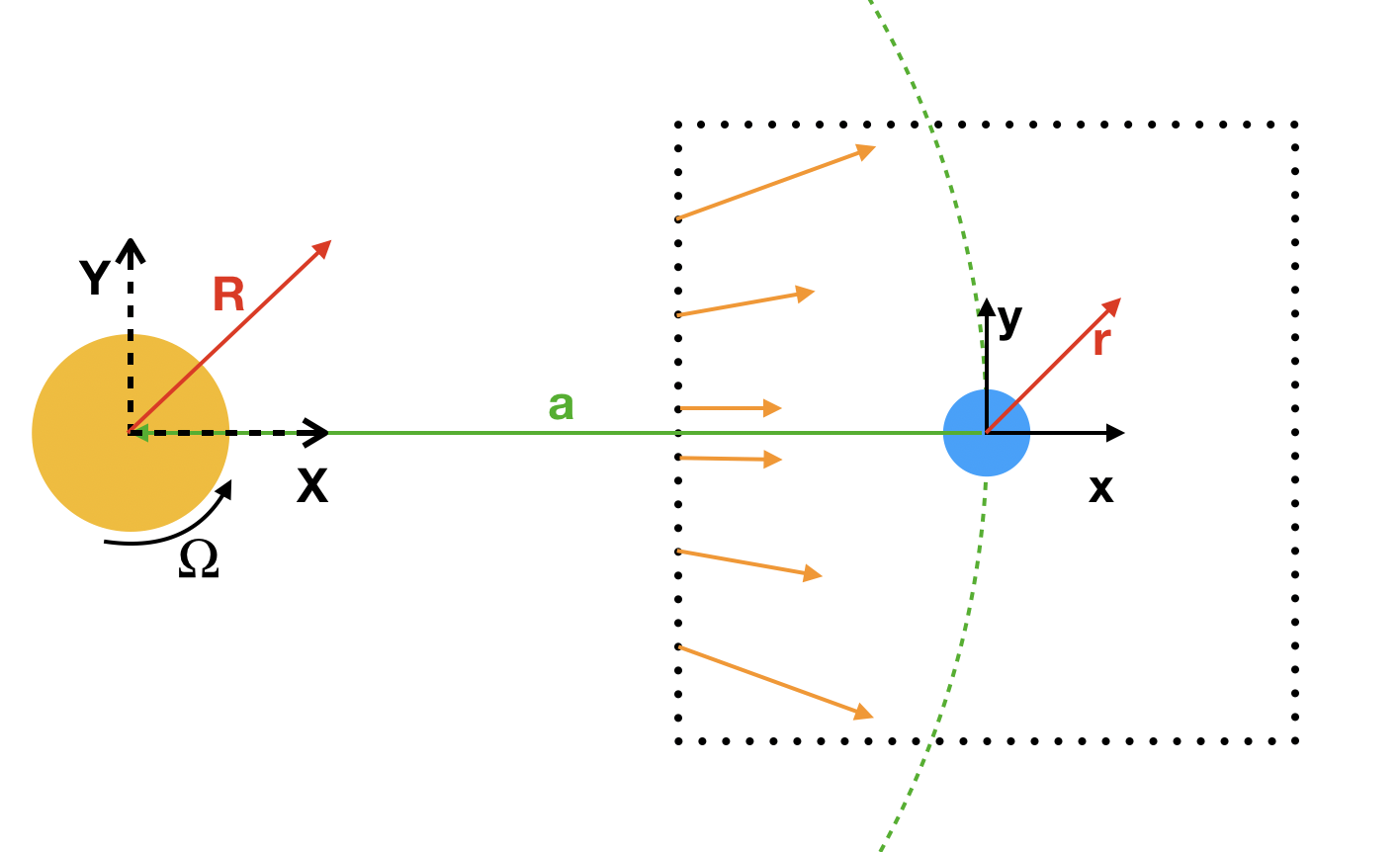}
    \caption{The coordinate system used in our simulations (not to scale). The orange circle represents the star, while the blue shows the planet. The uppercase letters show the direction of vectors in the stellar system, while the lowercase shows the axes in the planetary system. The black dotted square represents our computational grid, while the green dashed line shows the path of the planet's orbit. The orange arrows represent the changing direction and magnitude of the stellar wind velocity as it enters the simulation grid. Note that, because of the close distance to the star, we can not assume that the stellar wind has a plane-parallel injection, as is usually assumed in simulations of planets that orbit far from their host stars  \citep[e.g.][]{Carolan2019}.}
    \label{fig:coords} 
\end{figure}

Our \carito{3D} simulations are Cartesian and solve for 5 parameters in the corotating frame: the mass density ($\rho$), velocity ($u_x, u_y, u_z$) and thermal pressure ($P_T$). These are found through iteratively solving a set of ideal hydrodynamic equations that includes the mass conservation equation

\begin{equation}
    \frac{\partial \rho}{\partial t} + \nabla \cdot (\rho \vec{u}) = 0,
\end{equation}

\noindent the momentum conservation equation

\begin{equation}
    \frac{\partial(\rho\vec{u})}{\partial t} + \nabla \cdot [\rho \vec{u} \vec{u} + P_TI] = \rho\bigg( \vec{g} -\frac{GM_{*}}{(r-a)^2} \hat{R} - \vec{\Omega} \times (\vec{\Omega}\times\vec{R})-2(\vec{\Omega} \times \vec{u}) \bigg), 
\end{equation}

\noindent and energy conservation equation

\begin{equation}
    \frac{\partial \epsilon}{\partial t} + \nabla \cdot [\vec{u}(\epsilon + P_T )] = \rho \bigg( \vec{g} -\frac{GM_{*}}{(r-a)^2} \hat{R} - \vec{\Omega} \times (\vec{\Omega}\times\vec{R}) \bigg) \cdot \vec{u},
\end{equation}

\noindent where $I$ is the identity matrix, $\vec{g}$  the acceleration due to the planet's gravity, $G$  the gravitational constant, and $M_*$ is the mass of the star. $\vec{r}$ is the position vector relative to the planet, $\vec{a}$  the position of the star relative to the planet, $\vec{\Omega}$  the orbital rotation rate, and $\vec{R}$ is the position vector relative to the star, as shown in figure \ref{fig:coords}. The total energy density $\epsilon$ is given by: 
\begin{equation}
    \epsilon = \frac{\rho u^2}{2} + \frac{P_T}{\gamma -1}. 
\end{equation}
where the thermal pressure is $P_T = \rho k_B T/(\mu m_p)$, where $k_B$ is the Boltzmann constant,  $\mu$ is the  mean mass per particle and $m_p$ is the mass of the proton. Given that our calculations are done in the non-inertial frame, where the planet is fixed at the origin, we include the non-inertial forces in the hydrodynamic equations as source terms. These terms are shown in the right-hand side of Equations (3) and (4) and they are: the Coriolis force ($-2\rho(\vec{\Omega} \times \vec{u})$), the centrifugal force and stellar gravity that combined give rise to the `tidal force'  ($-\rho(GM_{*}/(r-a)^2) \hat{R} - \rho\vec{\Omega} \times (\vec{\Omega}\times\vec{R})$). 

\begin{figure}
    \centering
    \includegraphics[width=\columnwidth]{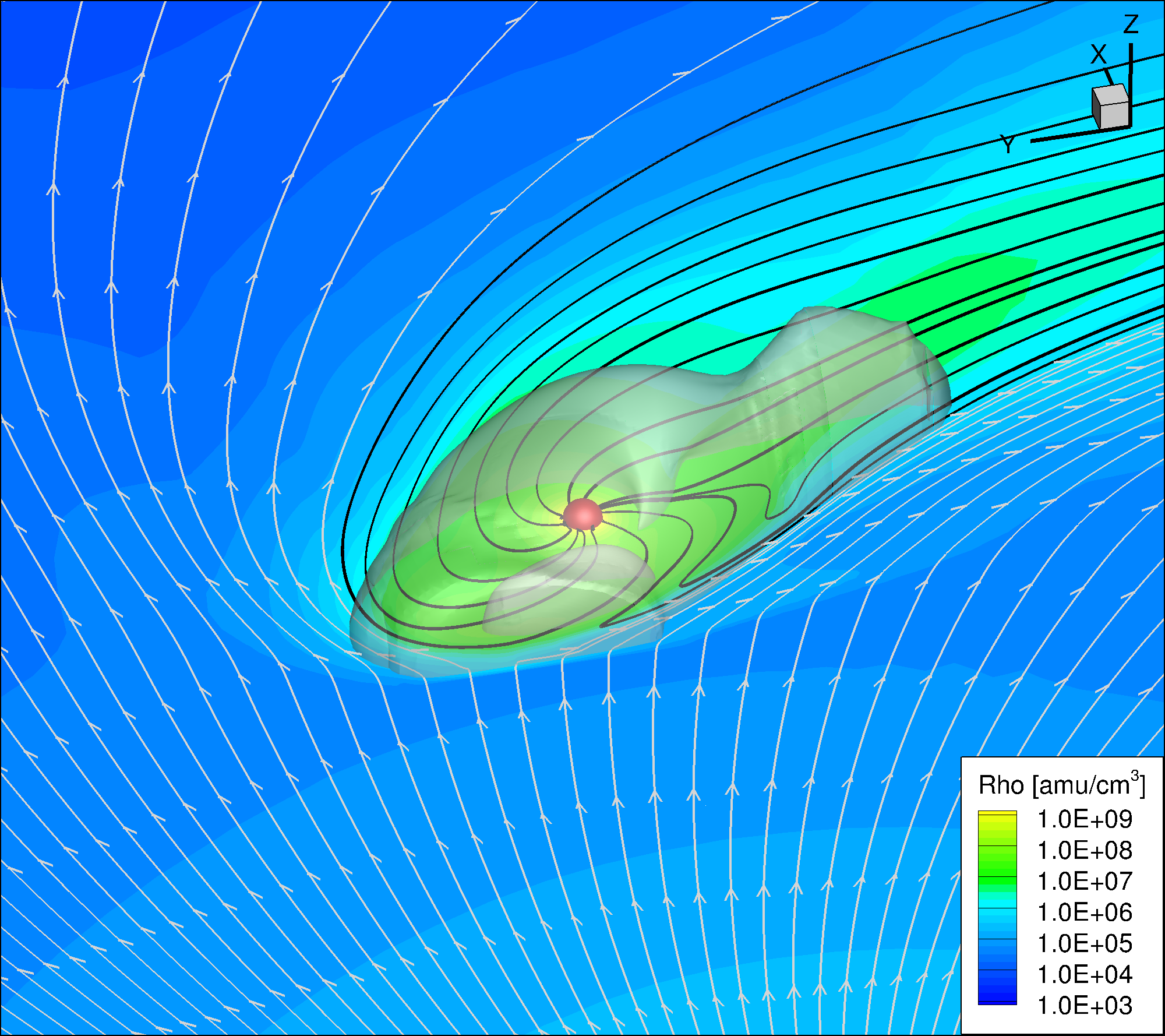}
    \caption{A 3D view of the HJ model for a stellar mass-loss rate of $10~\dot{M}_\odot$. The colour shows the density in the orbital plane. The gray surface marks the sonic surface around the planet. The grey streamlines show the flow of the stellar wind in the grid, while the black lines trace the escaping atmosphere of the planet.}
    \label{fig:3dfigure}
\end{figure}

Our simulations model the isothermal flow of ionised hydrogen around the planet, similar to the work of \citet{Bisikalo2013, CN2017}. To achieve this, we set the mean mass per particle $\mu=0.5~m_p$ and the polytropic index $\gamma\simeq 1$, which ensures a constant temperature in the outflow. The exoplanet is placed at the origin of a rectangular grid ($x, y =[-50, 50~r_p]$, $z=[-32, 32~r_p]$, where $r_p$ is the radius of the planet) as seen in figures \ref{fig:coords} and \ref{fig:3dfigure}. For simplicity we chose the planet to be tidally locked to the star, such that the star is always located at negative $x$, while the planet orbits in the positive $y$ direction. $z$ constructs the right-handed system. This approach is justified, given that 
many close-in exoplanets are thought to be tidally locked to their host star \citep{Kasting1993, Edson2011}. Our simulations contain 16 million cells and have a \blue{ minimum cell size} of $1/16~r_p$ within a radius of $5~r_p$, which gradually decreases towards the edge of the grid. We found increasing the maximum resolution of these simulations showed no significant change in the results. 

The inner boundary is placed at the surface of the planet ($r= 1~r_p$), where we keep the base density $n_{0,p}$ and temperature $T_p$ of the escaping atmosphere fixed throughout the simulations.  The base values adopted for each of the modelled planets  are shown in table \ref{tab:planets}. These values give rise to an escape rate $\dot{m}_0$, \blue{which matches that found in the 1D model}. Finally, we assume the velocity of the outflow to be reflective in the non-inertial frame, effectively setting the inner velocity to $\simeq 0$.

The simulation box has six outer boundaries, one for each face of the rectangular grid.
The stellar wind is injected to the grid at the negative $x$ boundary. The physical properties of the stellar wind are found by modelling an isothermal 1D stellar wind model at a given temperature $T_\star$ \citep{Parker1958}. The mass-loss rate ($\dot{M}$) is a free parameter of the stellar wind model, that we vary in our study to change the strength of the incoming stellar wind. For a given $\dot{M}$ and $T_\star$, the 1D solution provides the velocity and density of the stellar wind at every point on the face of the negative $x$ boundary. We orient the stellar wind to flow radially away from the star, but as it is injected in the simulation domain, we account for the Coriolis velocity ($-\vec{\Omega} \times \vec{R}$) to convert to the planet's reference frame. 
On all remaining faces of our rectangular grid, we use an inflow limiting boundary condition, similar to \citet{McCann2019}. This means that the outer boundary applies normal floating boundary conditions to cells where the velocity is directed outward, but set the momenta to 0 if the velocity component is directed inward. This is necessary as the Coriolis force can bend the flow of material close to the outer boundaries, such that uncontrolled inflows are generated when normal floating outflow conditions are applied. This can cause numerical issues such that no steady state solution is attainable, especially in smaller computational grids. The inflow limiting condition prevent this, ensuring that only the stellar wind is being injected into our grid.

\section{Simulations of reduced atmospheric escape in close-in planets}
\label{sec:MassLoss}

Using this computational setup we model two exoplanets for a range of stellar wind conditions: a Hot Jupiter similar to HD209458b but orbiting a \blue{more active} solar-like star (labelled ``HJ"); and a Warm Neptune similar to GJ3470b (labelled ``WN") which orbits an M dwarf. The relevant parameters used in the HJ and WN models are also provided in table \ref{tab:planets}.

\begin{figure*}
    \centering
    \includegraphics[width=0.9\textwidth]{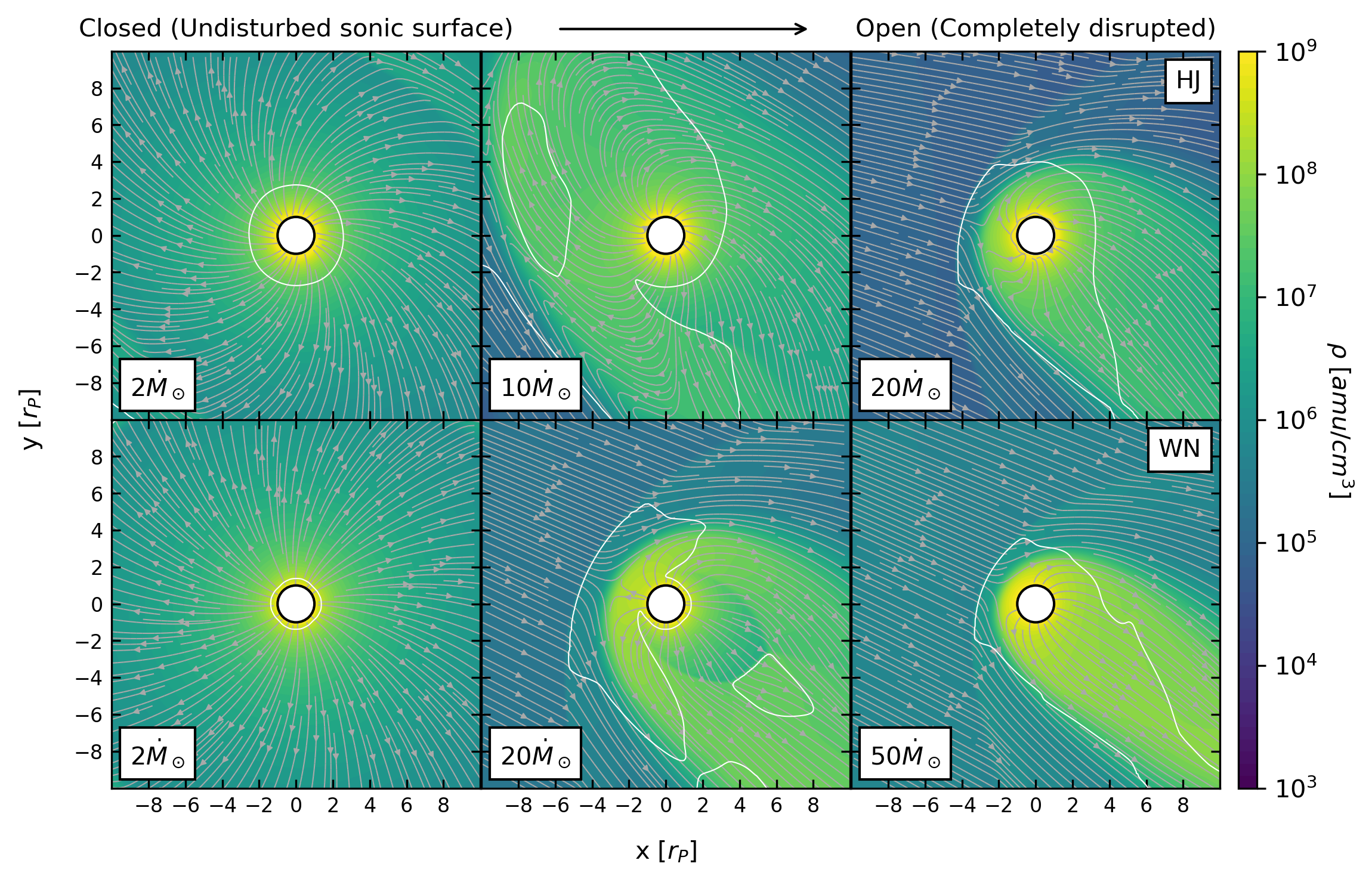}
    \includegraphics[width=0.9\textwidth]{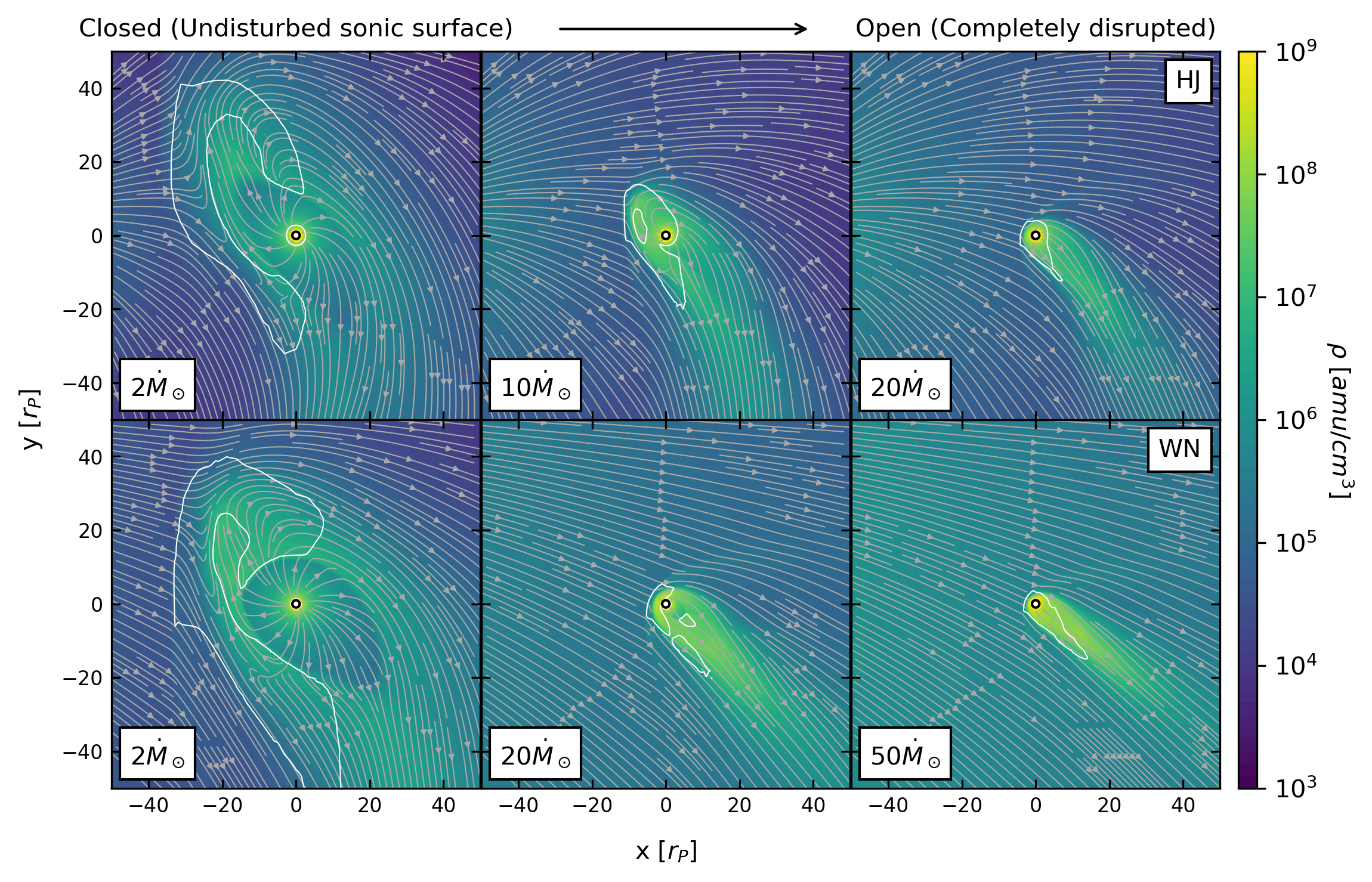}
    \caption{\blue{Top: }The orbital slice of 3 Hot Jupiter (HJ, top) and 3 Warm Neptune (WN, bottom) models, chosen to illustrate transition from ``closed" to ``open" sonic surfaces for each planet. \blue{``Closed'' refers to the scenarios where the inner (circular) sonic surface around the planet is not disturbed by the stellar wind. Contrary to that, ``open'' refers to the scenarios where this surface is disrupted.} The planet is orbiting in the positive y direction, with the star at negative x. The colour shows the distribution of density around the planets, while the white contour marks a Mach number of 1, which shows the sonic surface around each planet. The streamlines trace the velocity of material in each model, in the planet's  reference frame. Bottom: A zoomed out version of this figure, showing the orbital plane from our grids. \bluetwo{Here we can more clearly see the injection of the stellar wind from the negative x side of our grid, as well as the extension of material both ahead and behind the planet's orbit.}}
    \label{fig:models}
\end{figure*}

For each planet we compute 13 and 14 models respectively, increasing the stellar wind mass-loss rate from 0 to 100 times the solar mass-loss rate ($\dot{M}_\odot=2\times10^{-14}~M_\odot$/yr). Given that it is difficult to measure the winds of cool dwarf stars, we chose this large range of stellar mass-loss rates to establish if and when neglecting the presence of the stellar wind as a contributing factor to \blue{signatures of} atmospheric escape is appropriate. We note that this can only be achieved in multi-dimensional studies, as 1D models cannot account for the presence of a stellar wind. We chose to keep the temperatures of the isothermal stellar wind constant in each set ($T_{\star, {\rm HJ}} = 2\times10^6~K$, $T_{\star,{\rm WN}} = 1\times10^6~K$), retaining the same stellar wind velocity structure within each set of models. Varying the mass-loss rate, therefore, solely changes the stellar wind density. This setup allows us to investigate the effect that gradually increasing the stellar wind ram pressure \blue{($\propto \dot{M}$ $u_{\rm local}$, where $\dot{M}$ is the stellar wind mass loss rate, and $u_{\rm local}$ is the velocity of the stellar wind at the planet's orbit)} has on the planet's escaping atmosphere. The density structure and velocity flow in the orbital plane of three models from each set are shown in figure \ref{fig:models}. As we increase the stellar wind mass-loss rate we see the transition from type 2 / ``weak" to the type 1 / ``strong" scenarios shown by \citet{Matsakos2015} / \citet{McCann2019} respectively.

In the left panels of figure \ref{fig:models} we can see a large sonic surface created when the super-sonic stellar wind is shocked as it meets the super-sonic planetary outflow. As the ram pressure of the stellar wind increases this interaction happens closer to the planet. This confinement eventually affects the sonic surface of the escaping atmosphere, as seen in the changing white contour close to the planet from left to right panels in figure \ref{fig:models}. The inner \red{circular} sonic surfaces in the left panels are unaffected (labelled ``closed"). In the right panels the stellar wind has confined the outflow such that the sonic surface has been altered on all sides of the planet (labelled ``open"). The middle panels show the transition between these two states, where the day-side sonic surface has changed, while the night-side surface remains largely unaffected.  
\footnote{\red{The additional M=1 transition, in figure 5 is due to the fact that both the} \red{stellar and planetary winds are supersonic. As a result there are two shocks, one outer “bow shock” where the stellar wind is shocked, and an inner “termination shock” where the supersonic escaping atmosphere is shocked. This is similar to what is shown in models of the interaction between the heliosphere and the ISM \citep[eg.][]
{Zank2003}, where instead it is the inner stellar wind and outer ISM forming two shockwaves.}}
\blue{Note that the inner sonic surface of the WN planet is at $1.1~r_p$. Published models \red{(eg. Villarreal D'Angelo submitted)} of GJ436b, also a \carito{Warm Neptune}, have shown that the $M=1$ transition occurs farther out, where $M$ is the mach number. The low position of the sonic point in our 3D models is due to our choice of temperature (see figure \ref{fig:model_comp}). We will come back to how a larger sonic surface would change our results when we further discuss the limitations of our model in section \ref{sec:limitations}.}

\begin{table}
    \centering
    \caption{Summary of our simulation results. The stellar wind mass-loss rate ($\dot{M}$) is varied in each simulation, affecting the planetary atmospheric escape rate $\dot{m}$ and Ly-$\alpha$ absorption at mid-transit computed in the \blue{blue [-300 to -40 km/s] plus red [40 to 300 km/s]} wings \bluetwo{(including the broad-band absorption due to the planetary disc)}. Here, the superscripts HJ and WN refer to the \carito{Hot Jupiter}  and \carito{Warm Neptune}cases, respectively.}
    \begin{tabular}{cccccc}
        \hline
        $\dot{M}$ (HJ) & $\dot{m}^{\rm HJ}$ & $\Delta F ^{\rm HJ}$ &  $\dot{M}$ (WN) &  $\dot{m}^{\rm WN}$ &  $\Delta F ^{\rm WN}$\\
         $(\dot{M}_\odot$) & ($10^{10}$ g/s) & (\%) &$(\dot{M}_\odot$/yr)&($10^{10}$ g/s)&(\%)\\
        \hline
        0 & 58&20.7     &0  &6.5&8.8\\
        2 & 56&12.0     &1  &6.6&7.3\\
        4 & 55&8.6     &2  &6.4&5.3\\
        6 & 55&6.9     &4  &6.5&4.0\\
        8 & 57&6.4     &10 &6.4&3.7\\
        10 &51&6.0     &15 &6.4&3.7\\
        12 &40&5.9     &20 &6.5&3.7\\
        14 &37&5.8     &25 &6.4&3.7\\
        16 &35&5.8     &30 &6.2&3.6\\
        20 &30&5.5     &35 &6.0&3.5\\
        30 &27&5.1     &40 &5.8&3.4\\
        60 &22&4.2     &50 &5.4&3.2\\
        100&21&3.9      &75 &4.6&2.8\\
        -  &-   &-      &100&3.9&2.4\\
        \hline
    \end{tabular}
    \label{tab:planets2}
\end{table}

\begin{figure}
    \centering
    \includegraphics[width=\columnwidth]{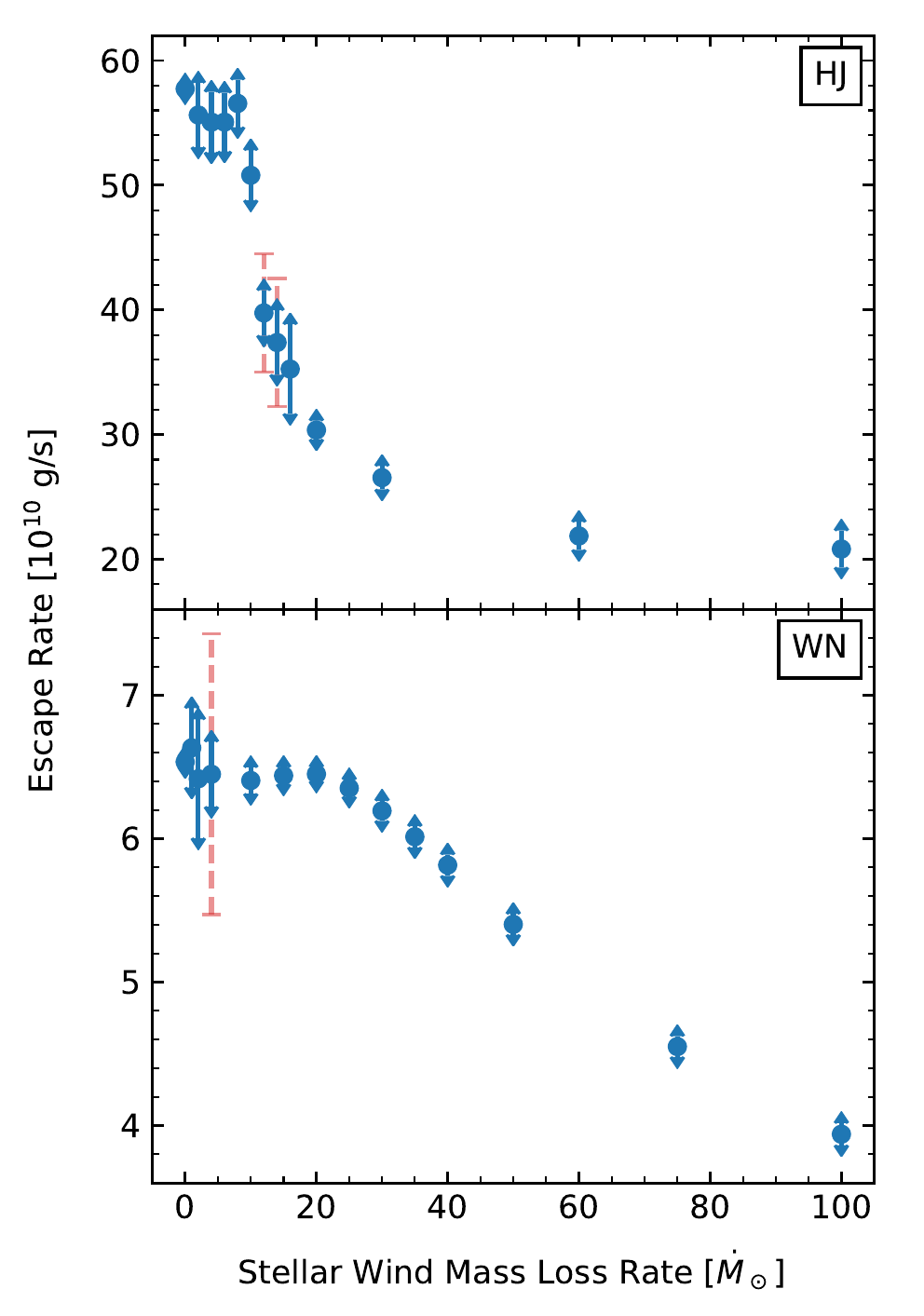}
    \caption{The variation of atmospheric escape rate with stellar mass-loss rate for each of our models. The points mark the mean escape rate, while the blue error bars illustrate one standard deviation from the mean. As some models reach a quasi-steady state solution, the average over several time steps is taken, and plotted in blue. For these models, the standard deviation at a single time step is shown in red.  }
    \label{fig:massloss}
\end{figure}

Previous models have shown that stellar wind confinement can affect the escape rate of the planets atmosphere \citep{Vidotto2020, Christie2016}. To investigate the difference in escape rate between our closed and open 3D models, we integrate the mass flux through concentric spheres around \carito{the} planet, obtaining the atmospheric escape rate:
\begin{equation}
    \dot{m} = \oint_A  \rho \vec{u} \cdot dA .
\end{equation}
Table \ref{tab:planets2} shows a summary of the results of our simulations. The escape rate from each model is plotted in figure \ref{fig:massloss}. Due to resolution, the escape rate can show small variations with distance from the planet, especially at the point where the stellar wind and escaping atmosphere meet. This small variation is quantified by the blue error bars in figure \ref{fig:massloss}, which show one standard deviation from the mean escape rate in each model. Some models only reach a quasi-steady state solution, showing small periodic variability with increasing time step. For these models the maximum deviation is shown as a red error bar, while the average deviation over one period of the quasi-steady state variability is shown in blue. In both model sets, increasing the stellar wind mass-loss rate has reduced the atmospheric escape rate. For the HJ models, the escape rate has been reduced by 65\% (from $5.8\times10^{11}$ g/s to $2.1\times10^{11}$ g/s, see table \ref{tab:planets2}). The WN models show a more gradual change, with a maximum reduction of 40\% (from $6.5\times10^{10}$ g/s to $3.9\times10^{10}$ g/s) over the range of stellar winds examined. 

As the stellar wind mass-loss rate increases, the escaping atmosphere is confined to a reduced volume around the planet, which is clearly visible from the velocity streamlines in figure \ref{fig:models} (left to right). This decelerates material on the day-side of the planet, redirecting it towards the planetary tail as seen in figure \ref{fig:models}. As the flow of the escaping atmosphere is further confined, this deceleration occurs closer to the planet, where it eventually inhibits the dayside flow from reaching super-sonic speeds. As a result, the atmospheric escape rate is gradually reduced as the flow's sonic surface is further disrupted. 

This change in escape rate occurs more suddenly in the HJ models than in the WN models as $\dot{M}$ is increased. We propose this is related to the \blue{ distance of the sonic point}. In the closed HJ and WN models these are approximately $2.5~r_p$ and $1.1~r_p$ respectively. As a result the outflow in the WN models must be confined relatively closer to the planet in order for the sonic surface to be affected. This results in a more gradual reduction of atmospheric escape in the WN models, as a stronger stellar wind is required to further confine this outflow and open the sonic surface. In the HJ models, this is not the case. The change from closed to open geometries happens over a much shorter range of $\dot{M}$, as the sonic surface is relatively further from the planet, and so can be more easily accessed by the stellar wind.

\section{Synthetic observations: reduced Lyman-$\alpha$ transit depth by stellar wind confinement}
\label{sec:Raytrace}

We use a ray tracing model to simulate the Ly-$\alpha$ line profile of each planet at mid transit. A full description of this model is outlined in \citet{Vidotto2018, Allan2019}. We further adapt this model to take our 3D grids as input. Given that the ray tracing model is constructed for an equally-spaced grid and calculations are done in the inertial reference frame, we interpolate our non-uniform 3D grid to contain 201 points equally-spaced in each dimension, describing the density, temperature, and line of sight velocity of the planetary material \red{in the (observer's) inertial frame}. Given that the stellar wind and the planetary outflow have very different temperatures, we separate planetary from stellar wind material by means of a temperature cutoff. To ensure we capture all the material escaping the planet, we use a temperature cutoff that is slightly higher than the temperature of the planetary outflow, ensuring that all the planetary material capable of absorbing in Ly-$\alpha$ is considered when producing the synthetic observations.
Our current \blue{3D} model does not treat \blue{both} the neutral and ionised portions of the planetary outflow,  \blue{it simulates purely the ionised part ($\mu=0.5$)}. Therefore, we estimate the neutral hydrogen density ($n_n$) from the ionisation balance equation solved self-consistently in the 1D escape model from \citet{Allan2019}, \blue{which tracks both the neutral and ionised density, as outlined in section \ref{sec_1dcode}}. Our 3D model yields the density of ionised hydrogen ($n_p$) which can be used with \carito{the}ionisation fraction \carito{from the 1D model} ($f_{\rm ion}$) to find the neutral density as follows:
\begin{equation}
    n_n = n_p \frac{1-f_{\rm ion}}{f_{\rm ion}}.
\end{equation}

\noindent Although this is not the most precise approach to calculate the neutral hydrogen density, this post-processing technique is a work-around adopted when the ionisation balance equation is not solved self-consistently with the hydrodynamics equations (similar approach has been used by \citet{Oklopcic2018} when calculating the population levels of helium, and by \citet{Lampon2020} when modelling helium in HD209458b's atmosphere). One limitation of this approach is that the 1D model computes the ionisation fraction along the star-planet line. Therefore, when we incorporate the ionisation fraction predicted in the 1D models in our 3D grid, the density of neutral material in the night side, or planetary tail, is not properly calculated. However, given that the inner regions  of our 3D simulations are approximately spherically symmetric and contain most of the absorbing material, using the resultant ion fraction from the 1D model is an acceptable approximation for computing the synthetic observations.

Once the neutral density is estimated,  the frequency $\nu$ dependent optical depth along the line of sight is given by:

\begin{equation}
    \tau_\nu = \int n_n \sigma \phi_\nu dx,
\end{equation}

\noindent where the observer is placed at positive x and $\phi_{\nu}$ is the Voigt line profile function. The absorption cross section  at line centre is $\sigma = {\pi e^2 f}/({m_e c})$, where $f = 0.416410$ is the oscillator strength for Ly-$\alpha$, $m_e$ is the mass of the electron, $e$ is the electron charge and $c$ is the speed of light. Using these, the fraction of transmitted intensity is given by:

\begin{equation}
    \frac{I_\nu}{I_*} = e^{-\tau_\nu}.
\end{equation}

\noindent Therefore, $1-{I_\nu}/{I_*}$ represents the fraction of intensity that is absorbed by the planet's disc and atmosphere. We shoot $201 \times 201$ stellar rays through the grid. Integrating over all rays, and dividing by the flux of the star allows for the frequency-dependent transit depth ($\Delta F_\nu$) to be calculated:
\begin{equation}\label{eqn_depth}
    \Delta F_\nu = \frac{\int\int (1-e^{-\tau_\nu}dydz)}{\pi R_*^2}.
\end{equation}

\begin{figure*}
    \centering
    \includegraphics[width=\textwidth]{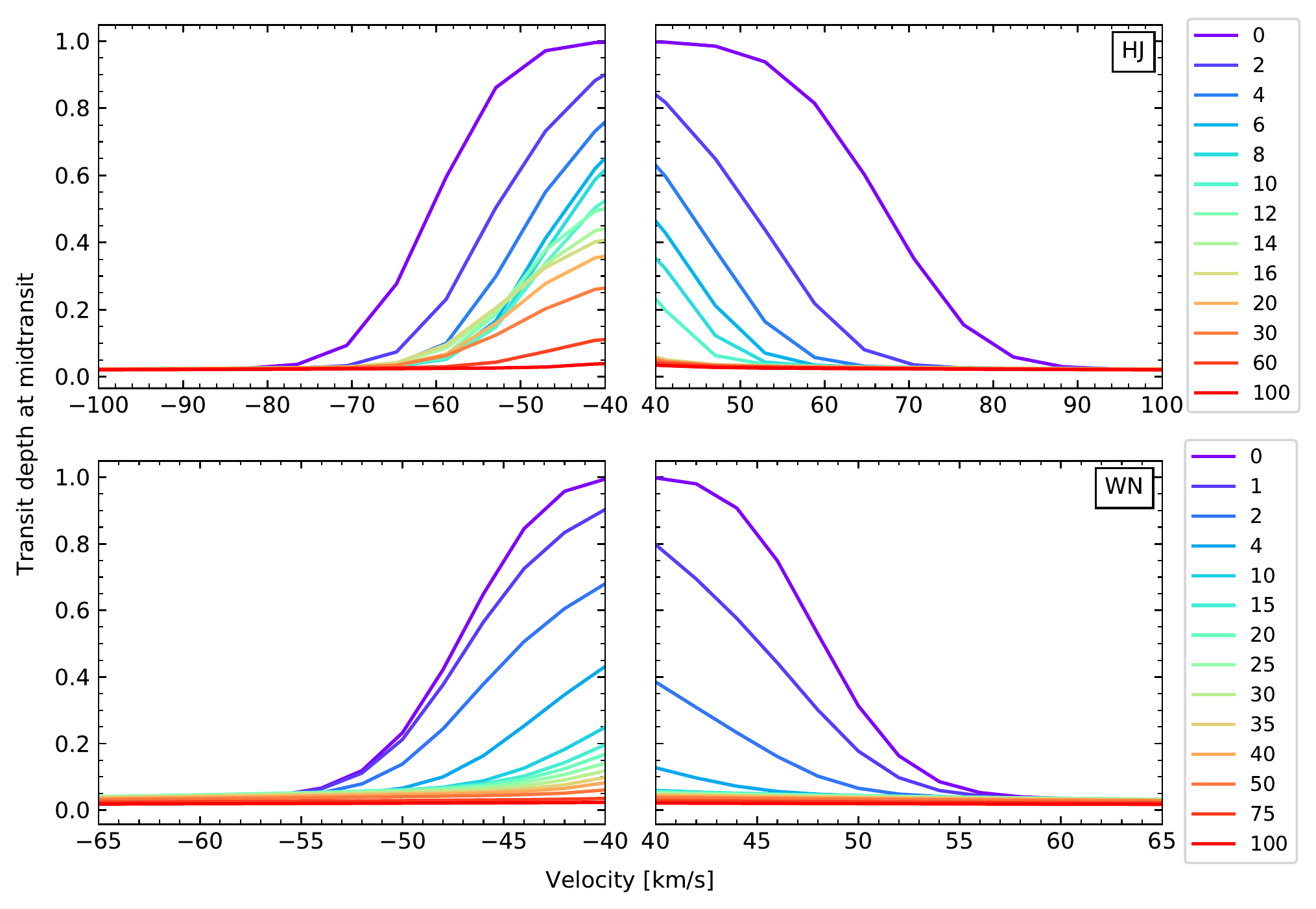}
    \caption{The blue (left) and red (right) wings of the \carito{in-transit Ly$\alpha$} line profile (\blue{equation \ref{eqn_depth}}) for each of our HJ (top) and WN (bottom) models. The colour shows the variation of stellar mass-loss rate from 0 to 100$~\dot{M}_\odot$. As the line centre is contaminated by interstellar absorption and geocoronal emission, we ignore the velocity range of -40 to 40 km/s. Note how the transit absorption is larger in the red wing for low stellar $\dot{M}$ and larger in the blue wing for high stellar $\dot{M}$.}
    \label{fig:raytrace}
\end{figure*}

For each of the models in the HJ and WN set, the transit depth as a function of velocity is shown in figure \ref{fig:raytrace}, \bluetwo{and the integrated percentage absorption is given in table \ref{tab:planets2}}.  
As the Ly-$\alpha$ line centre is dominated by interstellar absorption and geocoronal emission, we omit the line centre $[-40, 40 ]$ km/s from these plots. In both model sets, the maximum transit depth decreases as the planetary outflow is further confined. This is due to the volume of absorbing planetary material around the planet decreasing as the escaping atmosphere is confined. 

For most models, there is an obvious asymmetry in the line profile, with the blue wing showing more absorption than the red. In the low $\dot{M}$ models, there is significant dayside (redshifted) outflow towards the star, similar to that seen by \citet{Matsakos2015} in their type 2 interaction, and in the ``weak" stellar wind scenarios of \citet{McCann2019}. As $ \dot{M}$ increases, this dayside stream is suppressed completely, as seen in figure  \ref{fig:models}. What remains is the planetary tail, containing mostly blue shifted material (type 1 in \citet{Matsakos2015}; ``strong" stellar wind in \citet{McCann2019}), leading to the line profile asymmetry we see in figure  \ref{fig:raytrace}.

\begin{figure}
    \centering
    \includegraphics[width=\columnwidth]{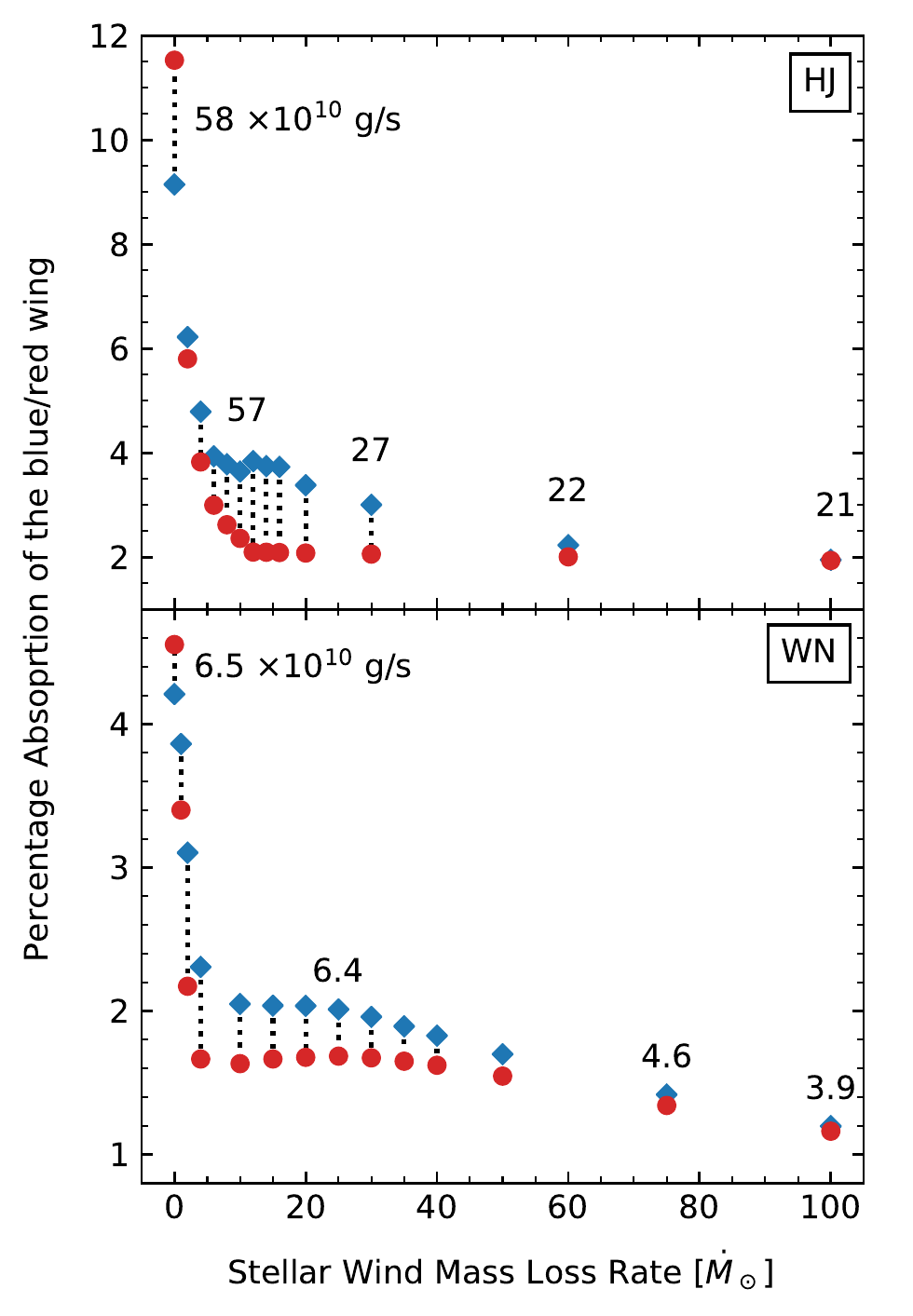}
    \caption{The percentage absorption in the blue (blue diamonds, -300 to -40 km/s) and red (red circles, 40 to 300 km/s) wings of Ly-$\alpha$ line as the stellar wind mass-loss rate is increased. 
    The dotted line links points from the same model. The top panel shows the results for the HJ model while the bottom shows the WN models. At various points the atmospheric escape rate of the planet is marked in units of $10^{10}$ g/s. }
    \label{fig:Integral2panel}
\end{figure}

\begin{figure}
    \centering
    \includegraphics[width=\columnwidth]{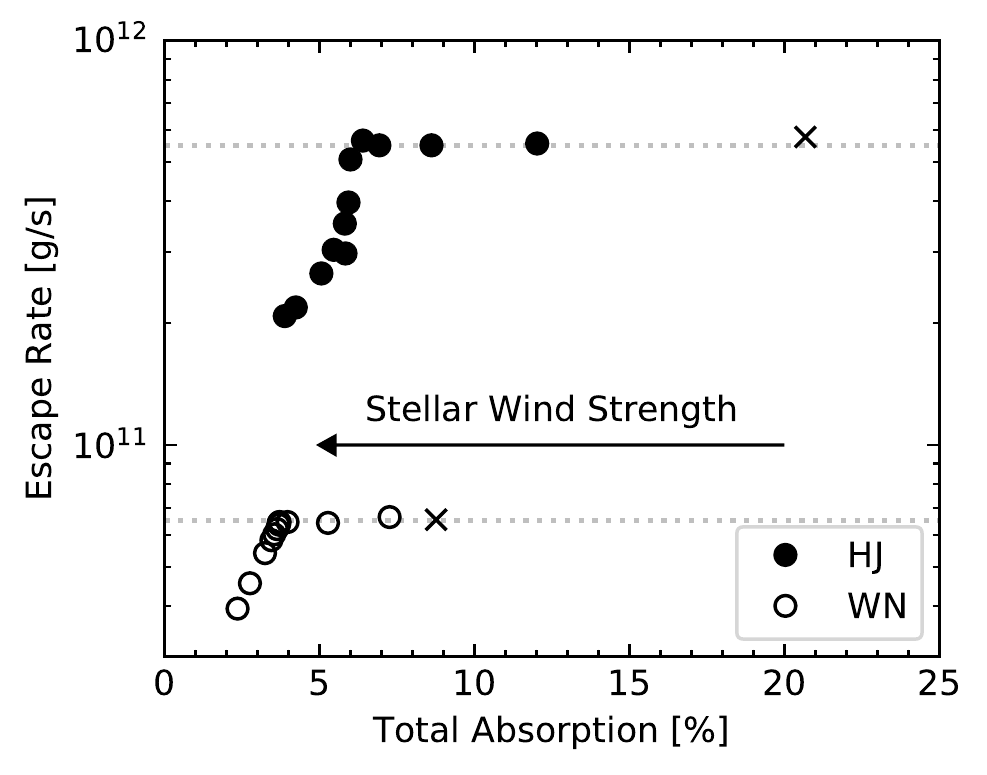}
    \caption{Atmospheric escape rate as a function of the percentage absorption of the blue$+$red wings of the Ly-$\alpha$ line at mid transit. The filled and open circles represent the HJ and WN models, respectively. For each set of models, we see a region where, although the escape rate is constant (marked with dotted lines), the transit absorption is not. The points marked with x are the models of each planet with no stellar wind, and from right to left, the stellar wind strength is increases (marked with an arrow).}
    \label{fig:integral}
\end{figure}

We quantify this asymmetry by integrating the modelled transit line profiles in figure \ref{fig:raytrace} over velocity in the blue \blue{[-300, -40 km/s] and red [40, 300 km/s] wings}. \blue{Note that our 3D model does not consider energetic neutral atoms (ENAs), and so the effect of these in the transit line profiles are not reflected in these calculations. For more discussion on this see section  \ref{sec:limitations}.} The results of this are shown in figure \ref{fig:Integral2panel}. 
Note that even in the case with no stellar wind the line shows asymmetry due to the orbital motion and tidal forces. We can clearly see that in both models at low stellar mass-loss rates the absorption in the red wing dominates due to the significant dayside stream as mentioned above. As the stellar mass-loss rate is increased, the absorption in the blue wing begins to dominate as this dayside flow is further confined. This is a more sudden change in the WN model than with the HJ planet, as the HJ planet has a stronger outflow/higher escape rate. For both models the level of asymmetry is approximately constant within a certain range of stellar mass-loss rates. For the HJ models, the blue wing absorbs 1-2\% more than the red from models between 10 and 30 $\dot{M}_\odot$. The WN models absorb roughly 0.25\% more in the blue between 10 and 40 $\dot{M}_\odot$. Above these ranges the difference in absorption between the blue and red wing gradually decreases  as the outflow becomes  significantly more confined. For stellar wind mass-loss rates $<10~\dot{M}_\odot$, despite the percentage absorption of each wing changing significantly (approximately 10\% in the the HJ, 3\% in WN) the atmospheric escape rate has not been affected as seen by the values on figure \ref{fig:Integral2panel}, and in figure \ref{fig:integral}. As mentioned in section \ref{sec:MassLoss}, the stellar wind further confines the escaping planetary material, which in turn shapes the absorbing material around the planet. However for low stellar mass-loss rates the sonic surface of the atmospheric outflow is not yet affected, thus the escape rate remains unchanged.

We further investigate the total absorption (blue + red) relationship with the atmospheric escape rate in figure \ref{fig:integral}. The filled and open circles represent the results for the HJ and WN models, respectively. The stellar wind mass-loss rate is increased from right to left points, with the right-most point in each set being the one without the presence of a stellar wind (marked with an x). For each set of models, we see a region of our parameter space where, although the planetary escape rate is constant (marked with dotted lines), the transit absorption is not. \blue{As mentioned previously, this is due to stellar wind confinement, whereby the increasing ram pressure of the stellar wind confines absorbing material closer to the planet, therefore covering a decreased region on the stellar disc \citep{Vidotto2020}.} In the HJ model set, a constant planetary escape rate of approximately $\dot{m}_{\rm HJ} \simeq 5.5\times10^{11}$ g/s produces 6\% to 21\% absorption, depending on the stellar wind condition. For the WN planet, a constant planetary escape rate of $\dot{m}_{\rm WN} \simeq 6.5\times10^{10}$ g/s produces 4\% to 9\% absorption. This reflects what is seen in figure \ref{fig:Integral2panel}, where the absorption of each wing has changed significantly, but the escape rate has not. As previously discussed in section \ref{sec:MassLoss}, in the models showing absorption within the quoted ranges, the sonic surface remains intact (our `closed' scenario), so the escape rate is unaffected by the changing stellar wind. Above a  mass-loss rate of 10 $\dot{M}_\odot$ (or 20 $\dot{M}_\odot$ for the WN model), i.e., below  an absorption of approximately 6\% for the HJ models (or 4\% for the WN models), the stellar wind begins to open the sonic surface, causing reduced atmospheric escape. In the range of WN models examined, the escape rate is reduced by approximately 40\%, from $6.5\times10^{10}$ to $4 \times10^{10}$ g/s, changing the total absorption by about 7\%. 
For comparison, a similar 40\% reduction in the HJ set, say, from $5.8\times10^{11}$ to $3.3\times10^{11}$  g/s, in escape rate occurs over a 15\% range in total absorption. 

\section{Discussion} 
\label{sec:Discussion}

\subsection{Implications for interpreting Ly-$\alpha$ signatures}
{The results presented in section \ref{sec:Raytrace} emphasise the importance of considering the stellar wind when interpreting transit observations. For example, suppose that a planet similar to our HJ model is observed with 5.5\% of absorption in the Ly-$\alpha$ line. In the presence of a stellar wind, we would interpret this observational result as being caused by a planetary outflow that has an intrinsic escape rate of $\sim 3\times10^{11}$ g/s (see Figure \ref{fig:integral}). However, had we not considered the presence of a stellar wind, we would have predicted a lower planetary escape rate to be able to reproduce this $5.5\%$ observed transit absorption. 
This is because, without stellar wind confinement, the absorbing material can now occupy a greater volume around the planet, yielding the same percentage absorption for a lower escape rate. Therefore, given the degeneracy between planetary escape rates and stellar wind strengths for a same percentage absorption, we suggest that,  in order to make a more accurate estimate of a planet's atmospheric escape rate, one should also consider the presence of a stellar wind.} 

Another consequence found in our study is  that the stellar wind confinement can act to mask observational signatures of atmospheric escape in Ly-$\alpha$. One instance of this could be with the \bluetwo{sub-Neptune} $\pi$ Men c. The surprising lack of detection of Ly-$\alpha$ absorption led \citet{Munoz2020} to suggest a lack of hydrogen in the atmosphere of $\pi$ Men c. If this is indeed the case, they proposed that further observations of heavier elements, such as OI and CII lines, could still reveal high atmospheric evaporation rate. 
An alternate explanation was proposed by \citet{Vidotto2020}, who suggested that the atmosphere of $\pi$ Men c could be hydrogen-rich, but would remain undetected if a significant reduction in escape was being cause by the stellar wind. They proposed that this could take place in $\pi$ Men c as they found that the stellar wind interaction  happened in the sub-sonic part of the planetary outflow. 

As we showed here, the position where the interaction happens depends on the strength of the stellar wind and on the properties of the escaping atmosphere. One important parameter in atmospheric modeling is the stellar high-energy flux that is deposited on the planet's atmosphere. In \citet{Vidotto2020}, they assumed a flux of $1340$ $\rm erg~cm^{-2}s^{-1}$ and a stellar wind mass-loss rate of $1\dot{M}_\odot$ for $\pi$ Men c. 
To investigate this further, \citet{Shaikhislamov2020} performed 3D simulations of the stellar wind of $\pi$ Men c and, using a similar stellar wind condition and high-energy flux than those assumed in \citet{Vidotto2020}, confirmed that the interaction was taking place in the sub-sonic regime, but did not obtain a significant reduction in the escape rate of the planet (only a factor of 2.5\% of the value obtained in their other simulations). We believe such a simulation performed by \citet{Shaikhislamov2020} falls in our partially-open scenario (middle panels in Figure \ref{fig:models}). 

Although we did not model the $\pi$ Men system, in the present work, we also found that the stellar wind might not necessarily lead to a significant reduction in escape rate, agreeing with \citet{Shaikhislamov2020}. Nevertheless, when looking at the whole parameter space covered in our simulations, we found reduction in evaporation rates of up to 65\% for the HJ simulations and 40\% for the WN simulations. However, more importantly, we also found that this reduction is not linearly related to the  depth of Ly-$\alpha$ transits -- even when the escape rate is not substantially reduced, the absorption signature of escaping atmosphere can be significantly affected. This is because the stellar wind confinement can change the volume of the absorbing material, without affecting too much the planetary escape rate. 

\subsection{Model limitations}\label{sec:limitations}

The aim of this work was to study in a systematic way how the stellar wind could affect atmospheric escape rates and how this would be observed in synthetic transits. Our model is relatively fast to run -- roughly, a simulation takes about 8h to compute in 40 processors, although it could take longer in some cases. This enabled us to model nearly 30 different physical setups. However, our model can still be improved. For example, our \blue{3D} simulations model the flow of ionised hydrogen. They are isothermal and ignore the effects of magnetic fields, radiation pressure and charge exchange. 

\blue{A caveat of the 3D isothermal model is that we cannot match both the ram pressure and sonic point found in the 1D model \footnote{\blue{It is a coincidence that the sonic point in the HJ simulation is at a similar distance in the 1D and 3D models}}. In this work, we chose to match the ram pressure of the escaping atmospheres, and as a result obtain sonic surfaces that are much closer to the planet \blue{than that predicted in the 1D model}. To investigate how matching the sonic point would affect our results, we run one additional model of the WN planet in a $100~\dot{M}_\odot$ stellar wind, ensuring that the sonic point remains at the same position as the 1D model, ie., at 1.9 $r_p$. This was done by reducing the temperature of the outflow from 5000 to 3100K, resulting in a  sonic point that is further out, and a lower velocity outflow. We chose a base density such that escape rate for this model with no stellar wind matches that of the previous version, $6.5 \times 10^{10}$ g/s. When the $100~\dot{M}_\odot$ stellar wind is injected into this model, we find that the atmospheric escape rate is reduced further than in the original, to $2.2 \times 10^{10}$ g/s (previously $3.9 \times 10^{10}$ g/s). This confirms that the larger the sonic point 
distance, the easier a stellar wind can disrupt this surface, and a greater reduction in escape rate will be found. However it also places importance on advancing our 3D model, as although we have now matched the sonic point, the ram pressure of the outflow is lower than found in the 1D model. This lower ram pressure provides less resistance against the incoming stellar wind, which when accompanied by the higher sonic point means the stellar wind can more easily disrupt the inner regions of the escaping atmosphere and affect the escape rate.}

\citet{CN2017} showed using 3D hydrodynamics simulations that the sonic surface of the escaping atmosphere is significantly different when assuming an isotropic versus an anisotropic temperature distribution at the planetary boundary (their anisotropic model mimics a nightside of the planet).
This obviously has implications on the atmospheric escape rate, with the anisotropic models of \citet{CN2017} showing  an escape rate that was approximately 50\% lower compared to the isotropic models. Temperature differences arise naturally in 3D models that solve for the radiative transfer of stellar photons through the planetary atmosphere \citep[e.g.][]{Debrecht2019}.
Although the day-night temperature differences are important to model, as they affect the planetary escape rate, we nevertheless expect that the hydrodynamic effects of the stellar wind on atmospheric escape is overall consistent to what we modeled here.

The presence of planetary magnetic fields, not included here, should also affect atmospheric escape.  
\citet{Owen2014}  demonstrated that magnetised exoplanets lose a factor of 4 to 8 times less mass than unmagnetised planets. This is because only a fraction of magnetic field lines around the planet remains open, along which ionised flows can escape. \citet{Khod2015} found that a dipolar magnetic field of 1G around a HD209458b-analog planet was capable of reducing mass loss by up to an order of magnitude. Additionally, in the case of a magnetised planet, the planetary magnetic field can deflect the stellar wind \citep[e.g.][]{Carolan2019}, which might no longer directly access the upper atmosphere of the planet. There has been recent debate in the literature discussing  whether a planet's magnetic field can prevent atmospheric erosion or could increase atmospheric losses \citep{2018MNRAS.481.5146B}. On one hand, the stellar wind is believed to erode atmospheres of unmagnetised planets, as is believed was the case of the young Mars, which was embedded in a much stronger young solar wind \citep{2007SSRv..129..207K}. On the other hand, the similar present-day ion escape rates of Mars, Venus and Earth have been used as counter-examples of planets that have very different magnetic field structures and yet present similar escape rates \citep{Strangeway2010}. There are still several unanswered questions oh how the magnetic field of a close-in exoplanet could affect atmospheric escape and transit signatures.

\blue{Finally, \blue{our model neglects two physical processes, namely radiation pressure and charge exchange, both of which have been suggested to produce the population of neutral atoms that absorbs in the wing of Ly$-\alpha$ at high blue-shifted velocities ($\sim 100$ km/s)}. Radiation pressure has been proposed to cause the acceleration of neutral hydrogen to speeds required to reproduce what is seen in Ly-$\alpha$ transits \citep{Bourrier2015, Schneiter2016}. However,  radiation pressure alone might not be as significant \citep{Debrecht2020}. Another alternative is that charge exchange processes could  produce high-velocity neutral hydrogen \citep{2008Natur.451..970H, 2014A&A...562A.116K, Bourrier2016, Tremblin2013, Shaikhislamov2016}, with a combination of charge exchange and radiation pressure processes best explaining observations of HD209468b \citep{Esquivel2019}. Though both of these effects can yield more high velocity hydrogen and thus affect the observational characteristics of the system, they are not thought to affect the inner regions of the escaping atmosphere. Charge exchange happens mostly around the shock,  as one needs stellar wind protons interacting with neutrals from the planet's atmosphere \citep{Shaikhislamov2016}. Radiation pressure could affect the outer layers of the planetary atmosphere, but gets absorbed as they penetrate the atmosphere (self-shielding). Due to this they are unlikely to significantly change the hydrodynamics of the planetary outflow though may contribute to the obtained transit line profile \citep{2018MNRAS.475..605C}.
For these reasons, we do not expect them to contribute to the disruption of the sonic surface that we study with our models.
}

\section{Conclusions}
\label{sec:Conclusion}
We have seen significant progress in modeling atmospheric escape in close-in planets over the past two decades \citep{2003ApJ...598L.121L, MurrayClay2009,Bourrier2013, Carolina2018, Debrecht2020}. However, many current models still neglect the interaction of the planet's upper atmosphere with the wind of the host stars. In particular, 1D hydrodynamic escape models are unable to treat the presence of the stellar wind --  for that, multi-dimensional simulations are required. It has been demonstrated that the wind of the host star can affect atmospheric escape in exoplanets \citep{Matsakos2015, Shaikhislamov2016, Carolina2018, McCann2019, Vidotto2020}. However, one of the open questions is currently for which systems the stellar wind would mostly affect atmospheric escape. One particular issue is that we do not have a full picture of how winds of cool dwarf stars vary from star to star. Cool dwarf stars are the most commonly known planet-hosts, but their winds are difficult to probe, with only a few techniques currently providing stellar wind measurements \citep[e.g.][]{Wood2005, VidottoDonati2017,2017A&A...599A.127F, 2019MNRAS.482.2853J}.  

In this work, we systematically examined the effects of stellar winds on  planetary atmospheric escape to determine whether and when neglecting the presence of the stellar wind is justified.  We used 3D hydrodynamic simulations to model the planetary outflowing atmosphere interacting with the stellar wind, which was injected in the simulation domain by means of an outer boundary condition. 
We performed this study on two different gas giant planets, a Hot Jupiter similar to HD209458b \blue{but orbiting a more active star} (HJ) and a Warm Neptune similar to GJ3470b (WN), and varied the stellar wind mass-loss rate from 0 to 100$\dot{M}_\odot$, where $\dot{M}_\odot=2\times 10^{-14}~M_\odot$/yr is the present-day solar mass-loss rate. In total, we performed nearly 30 3D hydrodynamics simulations of these systems.

For both planets, we found that as the stellar wind mass-loss rate was increased the planetary outflow was confined closer to the planet. As this confinement moved closer to the planet, planetary material could not properly accelerate, which eventually inhibited the dayside outflow from reaching super-sonic speeds. The inner regions of the escaping atmosphere were then affected, causing a reduced escape rate. For the HJ planet, the escape rate was reduced from $5.8\times10^{11}$ g/s, when no \carito{stellar} wind was considered, to $2.1\times10^{11}$ g/s, when a wind with 100$\dot{M}_\odot$ was considered. For the WN model, the planetary escape rates varied from $6.5\times10^{10}$ g/s (no stellar wind) to $3.9\times10^{10}$ g/s (strongest wind). 

This reduction happened more suddenly in the HJ models than in the WN models, which we proposed is related to the \blue{ distance of the sonic point.} 
As the stellar wind strength increases, the escaping atmosphere is confined closer to the planet. This eventually disturbs the sonic surface of the outflowing atmosphere, transitioning from a ``closed" to an "open" sonic surface configuration as seen in figure \ref{fig:models}. This is ultimately what reduces the escape rate of the planet's atmosphere, as the stellar wind now interacts with a subsonic planetary outflow and prevents an outflow from fully developing \citep{Christie2016, Vidotto2020}. As the sonic point in the HJ models is \blue{further from the planet}, the stellar wind can more easily affect it, so we see the escape rate of the HJ planet change over a short range in $\dot{M}$. In the WN planet, the sonic surface is very close to the planet. The stellar wind must now confine the escaping atmosphere much closer to the planet for the sonic surface to be affected, which results in a much more gradually transition from ``closed" to ``open" sonic surfaces in these models.

Using a ray-tracing technique, we investigated the possible observational signatures of this escape rate reduction in Ly-$\alpha$ transits. We found significant asymmetry towards the blue wing at mid-transit, which is to be expected when the day-side (redshifted) material is suppressed, and more material is funneled towards the planetary comet-like tail. This happens in the cases with larger stellar wind mass-loss rates. In the scenarios where only a weak wind (or no wind) were considered, the absorption in the red wing of the Ly-$\alpha$ line was larger, as some planetary material flows towards the star. 

We also found that the changes caused in the atmospheric escape rate by the stellar wind affects Ly-$\alpha$ transits in a non-linear way.
Across our set of models, the escape rate of the HJ planet was reduced from by 65\% with the Ly-$\alpha$ absorption changing  from 20.7 to 3.9\%.
For the WN planet, the escape rate was reduced by 40\%, with a corresponding change in Ly-$\alpha$ absorption from 8.8\% to 2.4\%.
However above 14\% absorption in the HJ set (5\% in the WN set) despite the percentage absorption changing significantly the atmospheric escape rate does not. These models represent the ``closed" sonic surface models, where despite the volume of absorbing material changing, the sonic surface remains unaffected and so the escape rate does no vary. We concluded  that the same atmospheric escape rate can therefore produce a range of absorptions depending on the strength of the stellar wind. 

Neglecting the stellar wind when interpreting Ly-$\alpha$ observations can also lead to under-estimations of the planet's atmospheric escape rate. An unconfined escaping atmosphere can occupy a larger volume around the planet, and so a lower escape rate is required to produce significant absorption. Contrary to this, if the escaping atmosphere is confined by a strong stellar wind, a higher escape rate can produce the same absorption as the unconfined scenario, as the density of absorbing material increases. These degeneracies emphasise the importance of considering the stellar wind when interpreting transmission spectroscopic transits, in order to accurately estimate the atmospheric escape rate.

\section*{Acknowledgements}
\blue{We thank the referee for their constructive review of this manuscript.} This project has received funding from the European Research Council (ERC) under the European Union's Horizon 2020 research and innovation programme (grant agreement No 817540, ASTROFLOW). The authors wish to acknowledge the SFI/HEA Irish Centre for High-End Computing (ICHEC) for the provision of computational facilities and support. This work used the BATS-R-US tools developed at the University of Michigan Center for Space Environment Modeling and made available through the NASA Community Coordinated Modeling Center. CVD acknowledge the funding from the Irish Research Council through the postdoctoral fellowship (Project ID: GOIPD/2018/659). 

\section*{Data Availability}
The data described in this article will be shared on reasonable request to the corresponding author.




\bibliographystyle{mnras}
\bibliography{mybib}


\label{lastpage}
\end{document}